\documentclass[12pt]{article}
\usepackage{amsmath}
\usepackage{amssymb}
\usepackage{amscd}
\usepackage{tabularx}
\usepackage{float}
\usepackage{graphics}
\usepackage{theorem}
\usepackage{enumerate}
\def\eqnum#1{}

\begin{document}

\title
{Earthquake Prediction: Probabilistic Aspect}

\date{}
\maketitle

\noindent {\footnotesize G.~MOLCHAN$^{a,b}$ and V.~KEILIS-BOROK$^{a,c}$} \\
\noindent{\footnotesize
$^{a}${\it International Institute of Earthquake Prediction Theory and
Mathematical Geophysics, \\
Russian Academy of Sciences, Moscow, Russia. E-mail: molchan@mitp.ru(GM) }} \\
{\footnotesize
$^{b}${\it The Abdus Salam International Centre
for Theoretical Physics, \\
SAND Group, Trieste, Italy }} \\
$^{c}${\footnotesize
{\it Institute of Geophysics and Planetary Physics and Department
of Earth \\
and Space Sciences, University of California, Los Angeles, USA
E-mail: vkb@ess.ucla.edu  }} \\

\bigskip

\noindent{\footnotesize Version: 31~Jan.~2008 }

\bigskip

\bigskip

\noindent{{\footnotesize \bf S~U~M~M~A~R~Y}} \\
\noindent{\footnotesize A theoretical analysis of the earthquake
prediction problem in space-time is presented. We find an explicit
structure of the optimal strategy and its relation to the
generalized error diagram. This study is a generalization of the
theoretical results for time prediction. The possibility and
simplicity of this extension is due to the choice of the class of
goal functions. We also discuss issues in forecasting versus
prediction, scaling laws versus predictability, and measure of
prediction efficiency at the research stage.}


\newpage

\section{Introduction}

The sequence of papers [Molchan 1991, 1997, 2002] was an attempt at
a probabilistic interpretation of what had been done in empirical
earthquake prediction during the 1980-1990s. These papers deal with
the problem of predicting the time of a large event in a fixed
region.

The
prediction involved the following concepts: {\it the information
flow} $I(t)$ used for prediction; {\it a prediction strategy} $\pi$
consisting of a sequence of decisions $\pi(t)$ that are relevant to
consecutive time intervals $(t, t+\Delta)$; {\it a decision}, which is
based on the information $I(t)$, and which is to choose an alarm level
for a time $\Delta$ (the zero level means an absence of alarm);
{\it losses}, which result from $\pi (t)$ and depend on whether the
decision is suitable for the actual seismic situation in $\Delta$;
{\it the goal of prediction}, which is to minimize a loss functional
for the monitoring period $T\gg 1$.

In the general case the optimal strategy is found as the solution of
a Bellman-type equation. However, there is one important case (at
least, at the research stage of prediction) for which the optimal
strategy is described explicitly, viz., the case where the goal
function can be described in terms of known prediction
characteristics: the rate of alarm time, $\tau$, and the rate of
failures-to-predict, $n$. The optimal strategy is then described
with the help of (a) conditional intensity of target earthquakes
given $I(t)$ and (b) the $n\,\& \,\tau$ (error) diagram, $\Gamma$
(Fig.~1). The latter is defined as the low bound of the set of the
prediction characteristics $(n,\tau)\in [0,1]^2$ that are relevant
to all possible strategies $\pi$ based on $I=\{ I(t)\}$.

If the flow $I$ is trivial, i.e., supplies no information for
prediction, then $\Gamma$ consists of the diagonal $D$ of the square
$[0,1]^2$: $n+\tau =1$. The curve $\Gamma$ is a decreasing convex
function. The greater the amount of information available, the
larger is the distance between curves $\Gamma$ and $D$. More
precisely, the condition $I_1(t) \subseteq I_2(t)$ implies
$\Gamma_1\ge \Gamma_2$. In the ideal case, $\Gamma$ degenerates to
the point $n=\tau =0$.

In actual practice the target earthquakes are large, hence rare,
events. This causes difficulties for statistical validation of a
prediction algorithm in a small region. That difficulty is being
overcome by parallel application of an algorithm in different
regions (e. g. algorithm $M8$ [Kossobokov and Shebalin, 2002] and
RTP algorithm [Keilis-Borok et al., 2004]). Prediction results are,
as before, presented using the error diagram, where $\tau$ is
replaced with the rate of space-time alarms $\tilde{\tau}$. The
properties of the modified diagram have not been studied yet.
Moreover, the generalization of $\tau$ itself is not unique. For
example, $\tilde{\tau}$ can be represented by the area of the alarm
space $A$ or by the expected number of target events within $A$,
i.e., $\lambda (A)$. Thus the case of space-time prediction needs
analysis, and such an analysis is presented below (see Section~3).

Next, we also discuss two more issues: the relation between
prediction and forecasting (sect.~4), and the relation between
predictability and self-similarity (sect.~5). These issues seem to
be urgent, considering that forecasting is dominant in prediction
research today, and the scaling laws indicating self-similarity are
frequently regarded as an obstacle in the way of predictability.

\section{Time Prediction}

Let us remind some facts concerning the simplest situation (see
below) in predicting the time of a target event in a fixed region
[Molchan, 2002].

The sequence of target events in the region will be considered as a
random stationary point process $dN(t)$, where $N(t)$ is the number
of events in the interval $(0, t)$ and $P(\Delta N(t)\ge 2)=o(\Delta
t)$. The prediction of $dN(t)$ is based on the information flow
$I(t)$, such that the $\{ dN(t), I(t)\}$ form a stationary ergodic
process; $I(t)$ may be thought of as a catalog of earthquakes in a
moving time interval $(t-t_0, t-t_1)$ with $t_0>t_1\ge 0$ fixed. A
prediction strategy $\pi =\{ \pi (t)\}$ consists of a sequence of
decisions $\pi (t)$: $\pi =1$ means an alarm during $(t, t+\Delta)$,
while $\pi =0$ means an absence of alarm. The occurrence of a target
earthquake during an alarm is termed a success. Each decision is
based on $I(t)$. The strategies are stationary and related in a
stationary manner to the process $\{ dN(t), I(t)\}$.

The following prediction results are to be recorded during time
$T=S\Delta$:

\begin{eqnarray}\label{qq1}
\tau_{_T} = S^{-1} \sum^S_{k=1}{\bf{1}}_{\{ \pi(t_k)=1\} },
\quad t_k=k\cdot \Delta
\label{qq1}
\end{eqnarray}
and

\begin{eqnarray}\label{qq2}
n_{_T} = S^{-1} \sum^S_{k=1}{\bf{1}}_{\{ \pi(t_k)=0\} }
{\bf{1}}_{\{ dN(t_k)=1\} } [S/N(T)].
\label{qq2}
\end{eqnarray}
where the logical function ${\bf{1}}_A$ equals 1 if $A$ is true and
0 otherwise. These statistics determine the empirical rates of alarm
time and failures-to-predict.

It follows from the above assumptions
that $\tau_{_T}$ and $n_{_T}$ have deterministic limits $\tau$ and $n$,
respectively,
as $T\to \infty$. They characterize the prediction capability of a strategy
$\pi$ based on the information $I=\{ I(t)\}$. On the other hand,
the $n\,\&\,\tau$
diagram mentioned in Introduction characterizes the
prediction capability of $I=\{ I(t)\}$.

Minimization of a goal function of type $\varphi (n,\tau)$, symbolically

\begin{eqnarray}\label{qq3}
\varphi (n,\tau) \Rightarrow \min\limits_{\pi},
\label{qq3}
\end{eqnarray}
is called here the {\it simplest prediction problem}. The choice of
$\varphi$ is governed by the particular applications of prediction
considered. There are only two general limitations: $\varphi$ should
increase with increasing $n$ and $\tau$ and the level sets $\{ n,
\tau : \varphi \le c\}$ should be convex.

Typical examples
of $\varphi$ that are used at the research stage are $\max (n,\tau)$
and $n+\tau$. The strategy that optimizes the first of these functions
is called the minimax strategy, for which $n=\tau$. The quantity
$e=1-(n+\tau)$ is frequently used to characterize
the efficiency of a prediction; it is the
higher the closer $e$ is to 1.
An example of $\varphi$ expressed in terms of damage is

\begin{eqnarray}\label{qq4}
\varphi = \alpha \lambda n + \beta \tau,
\label{qq4}
\end{eqnarray}
where $\lambda$ is the rate of target events, $\alpha$ is the cost
resulting from a failure-to-predict, $\beta \Delta$ is the cost of
maintaining an alarm during $(t, t+\Delta)$. Therefore, (\ref{qq4})
gives the loss rate entailed by $\pi$.

We now describe the structure of the optimal strategy. Let

\begin{eqnarray*}
r(t) = \lim_{\Delta \to 0}P\{ \Delta N(t)>0 \, \,\vert \, \,
I(t)\}/\Delta
\end{eqnarray*}
be the conditional rate of target events given $I(t)$. The optimal
strategy in the problem (\ref{qq3}) then declares an alarm every
time $r(t)$ exceeds a threshold $r_0$. The threshold is $r_0 =\beta
/\alpha$ when (\ref{qq4}) is used. In the general case of $\varphi
(n, \tau)$, we have to find the level $c$ such that the line  $\{
\varphi =c\}$ is tangent to the error diagram $\Gamma$ (see Fig.~1).
Suppose this occurs at a point $Q=(n_0,\tau_0)$. Then

\begin{eqnarray*}
r_0 = - \lambda \frac{dn}{d\tau} (Q) = - \lambda
\frac{\partial \varphi}{\partial \tau} \bigg /
\frac{\partial \varphi}{\partial n} (Q)
\end{eqnarray*}
where $dn/d\tau$ is the slope of $\Gamma$ at $Q$.

\begin{figure}[h]
\centering{\includegraphics{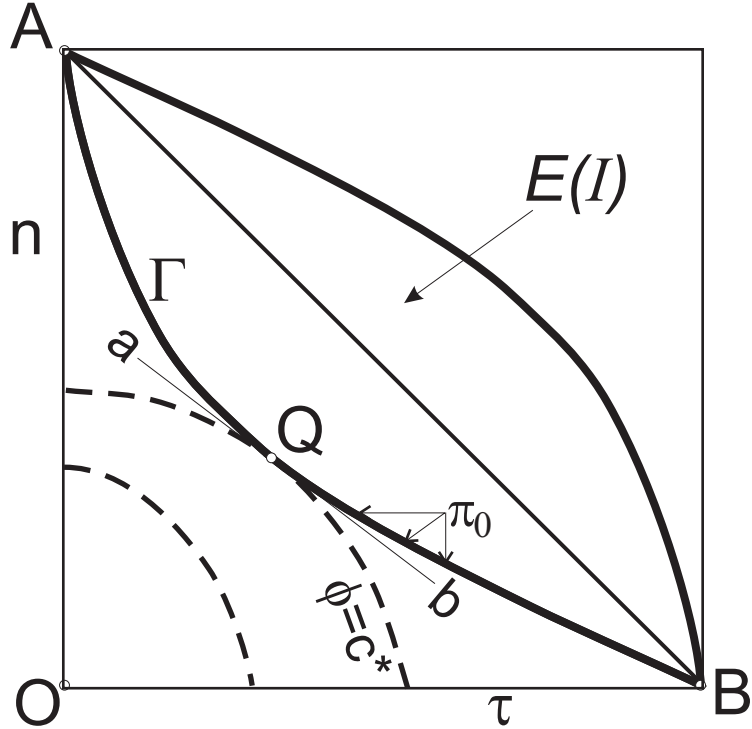}}
\caption{Error set ${\cal
{E}}(I)$ for prediction strategies based on a fixed type of
information $I=\{ I(t)\}$. The point $A$ corresponds to an
optimistic strategy, the point $B$ to a pessimistic strategy, the
diagonal $D=AB$ corresponds to strategies of random guess. $\Gamma$
is the error diagram of optimal strategies. {\it Small arrows}
indicate strategies better that $\pi_0$, i.e. strategies with  $n
\le n(\pi_0)$ and $\tau \le \tau(\pi_0)$. {\it Dashed lines} are
isolines of a loss function $\varphi (n,\tau)$; the isoline of level
$c^*$ is tangent to $\Gamma$ at the point $Q$, which corresponds to
the optimal errors in the problem (\ref{qq3}). The line $(a,b)$ is
tangent to $\Gamma$ at $Q$ and separates the two convex sets ${\cal
{E}} (I)$ and $\{ \varphi \le c^*\}$.}
\end{figure}

{\it The Relation to Hypothesis Testing}. We remind a classical
hypothesis testing problem in mathematical statistics (see, e.g.,
Lehmann, 1959). Consider an observation $\xi$, which may be a
scalar, a vector, or a functional object. It belongs to the
population with distribution $P_0(dx)$ (hypothesis $H_0$) or to the
population with distribution $P_1(dx)$ (hypothesis $H_1$). A
decision $\pi (\xi)=0$ or 1 in favor of $H_0$ or $H_1$,
respectively, entails errors of two kinds, viz.,

\begin{eqnarray*}
\alpha = P_0\{ \pi (\xi) = 1\} \quad \mbox {\rm and} \quad
\beta   = P_1\{ \pi (\xi) = 0\}.
\end{eqnarray*}
Let us fix $\alpha$ and minimize the error $\beta$ by a suitable
choice of $\pi$. The lemma of J.~Neyman and E.~Pearson reads that,
under certain regularity requirements, the optimal rule is such that
$\pi (\xi)=1$, as soon as

\begin{eqnarray*}
{\cal {L}} (\xi) = P_1(dx)/P_0(dx) \vert_{x=\xi}\ge c(\alpha),
\end{eqnarray*}
otherwise $\pi (\xi )=0$; note that the threshold depends on $\alpha$.

In applications the power
of the optimal test, $1-\beta$, is considered as a function of $\alpha$,
and called the Relative Operating
Characteristic (ROC), see [Swets, 1973].

The prediction problem (\ref{qq3}) is remarkable in that it can be
interpreted in terms of hypothesis testing, so that the
characteristics $(n, \tau)$ become errors of the two kinds [Molchan,
2002]. The crucial observation for this is the following: the
globally (in time) optimal strategy in (\ref{qq3}) consists of
locally optimal decisions on small time intervals $(t, t+\Delta)$.
One can therefore disregard the global prediction problem and
consider it on the interval $(t, t+\Delta)$. In this case $\pi (t)$
interprets incoming information $\xi =I(t)$ in terms of whether a
target event will or will not occur in the interval $(t, t+\Delta)$.
The characteristics $(n, \tau)$ become errors of the two kinds, if
$P_0$ is the natural probability measure for the data $I(t)$ at time
$t$, while $P_1(dx)$ is the conditional measure for $I(t)$ given
$dN(t)=1$.

Recalling the definition of the risk function $r(t)$, one has

\begin{eqnarray*}
\nonumber
P_1(dx)&=&P\{ dN(t)=1, I(t)\in dx\} / P\{ dN(t)=1\} \\
&=&P\{ dN(t)=1\, \vert \, I(t)=x\} P_0\{ I(t)\in dx\} / P\{ dN(t)=1\} \\
\nonumber
&=&r(t) P_0(dx)/\lambda.
\end{eqnarray*}
Hence $P_1(dx)/P_0(dx)=r(t)/\lambda$.
Furthermore, since  $n$ and $\tau$ are identical with the errors arising
from testing $H_1$ vs. $H_0$, we have

\begin{eqnarray*}
\mbox {\rm ROC} = \{ (1-n, \tau):(n,\tau) \in \Gamma \} = \Gamma^c
\end{eqnarray*}
that is, the curves ROC and $\Gamma$ are dual.

For this reason $\Gamma^c$ is sometimes called  ROC and sometimes
the Molchan diagram. However, these names have different
implications. The first name (ROC) always focuses our attention on
errors of two kinds in a statistical problem, while the names
$n\,\&\,\tau$ or error or Molchan diagram emphasize the connection
between two of the many characteristics of prediction. The ROC
interpretation of the curve $\Gamma$ is possible thanks to specific
features of the goal function and to the structure of the globally
optimal strategy. With a modified goal function, the error diagram
loses its relation to optimal strategies. The reason for this is
that locally optimal decisions do not generally constitute the
globally optimal strategy [Molchan \& Kagan, 1992].

In this context we mention the case of prediction for an
inhomogeneous Poisson process with a periodic rate function. It is
commonly thought that the prediction of a Poisson process is
trivial, and therefore does not deserve consideration. Molchan
[1997, 2002] showed that this is not true, if the losses also
include some cost for each switching from alarm to nonalarm and back
again. An optimization problem of this kind is reasonable to avoid
the cry wolf attitude.

Leaving aside the unimportant discussion of a suitable name for the
$n\,\&\,\tau$ diagram, we put new questions: what are the analogues
of $\Gamma$ and $D$ for  space-time prediction? What is the
structure of the optimal strategy for a goal function that is
similar to (\ref{qq3})?

\section{Space-time Prediction}

For a theoretical analysis of prediction of large events in
space-time it is sufficient to divide region $G$ into disjoint parts
$G_i$ and to consider the vector point process

\begin{eqnarray*}
dN(t) = \{ dN^{(1)}(t),\ldots, dN^{(k)}(t) \},
\end{eqnarray*}
where the component $dN^{(i)}(t)$, $P(\Delta N^{(i)}(t)\ge
2)=o(\Delta )$ describes the time sequence of target events in
subregion $G_i$. In that case a prediction strategy $\pi (t)=\{
\pi^1(t),\ldots, \pi^k(t)\}$ consists of the sequence of decisions

\begin{eqnarray*}
\pi^{(i)}(t)=
\begin{cases}
1                &\text{alarm in $G_i\times \Delta t$}\\
0                &\text{no alarm in $G_i\times \Delta t$,}
\end{cases}
\end{eqnarray*}
the decisions being based on the information $I(t)$.

Again the prediction results will be characterized by (\ref{qq1}) and
(\ref{qq2}). Here, $\tau_{_T}=(\tau_{_T}^1,\ldots,\tau_{_T}^k)$ is a vector
whose $i$-th component defines the ratio of
alarm time in $G_i$ during time $T$. When the vector process
$(dN(t), I(t), \pi (t))$ is ergodic and stationary, the numbers
$n_{_T}$ and $\tau_{_T}$ have the deterministic limits $n\in [0,1]$
and $\tau \in [0,1]^k$, respectively, as $T\to \infty$.

The use of all possible strategies $\pi$ based on $I=\{ I(t)\}$ yields
the error set $\{ (n, \tau )\} ={\cal {E}}$ as a subset of the cube
$[0,1]^{k+1}$.

The set ${\cal {E}}$ is convex. This can be demonstrated as follows.
Having two strategies, $\pi_1$ and $\pi_2$, with the characteristics
$(n,\tau)_i, i=1,2$, we can devise a new one with the errors
$(n_1p+n_2q, \tau_1p+\tau_2q)$, where $p+q=1$ and $0\le p \le 1$. To
do this, it is sufficient at every time step to use $\pi_1(t)$ and
$\pi_2(t)$ in a random manner, with probabilities $p$ and $q$,
respectively. Changing $p$ from 0 to 1, we get the straight segment
that belongs to ${\cal {E}}$ and connects $(n, \tau )_1$ with  $(n,
\tau )_2$. Therefore, ${\cal {E}}$ is convex.

The set ${\cal {E}}$  always contains the following simplex:

\begin{eqnarray}\label{qq5}
\tilde{D}: \quad n + <\lambda, \tau >/\Lambda = 1, \quad (n, \tau )
\in [0,1]^{k+1}, \quad \Lambda =\sum \lambda_i, \label{qq5}
\end{eqnarray}
where the $\lambda =(\lambda_1, \ldots, \lambda_k)$ are the rates of
target events in the subregions $\{ G_i\}$, and $<a,b>=\sum a_ib_i$.
Equation (\ref{qq5}) is satisfied by the following strategy based on
trivial information. Let us declare an alarm during $(t,t+\Delta )$
in subregion $G_i$ with probability $p_i$, $\sum p_i\le 1$. Then the
success rate in $G_i$ is $\lambda_i p_i/\Lambda$. Therefore, we have
$n=1-<\lambda, p>/\Lambda$ and $\tau_i =p_i$, i.e., (\ref{qq5})
becomes an identity.

The simplex (\ref{qq5}) is an analogue of the diagonal $D$ used in
the time prediction. The boundary of the convex set ${\cal {E}}$
that lies below the plane (\ref{qq5}) will be denoted
$\tilde{\Gamma}$ and termed the error diagram as above. We shall
show that the diagram defines optimal strategies. To do this, we
consider a function $\varphi (n,\tau)>0, \tau =(\tau^1, \ldots,
\tau^k)$ that is increasing with respect to each argument, and
require that the level sets  $\{ \varphi (n,\tau)\le c\}$ be convex
for any $c>0$. Now we define the goal of time-space prediction using
(\ref{qq3}) with  $\tau =(\tau^1,\ldots,\tau^k)$.

Denote $r^{(i)}(t)=\lim_{\Delta \to 0}P\{ \Delta N^{(i)}(t)=1 \,
\vert \, I(t)\}/\Delta$, the conditional rate of target events in
subregion $G_i$, given the information $I(t)$, and denote by
$\lambda_i$  the unconditional rate.

{\bf Statement~1.} {\it The optimal strategy for the space-time
prediction with the goal function} (\ref{qq3}) {\it declares an alarm in}
$G_i\times [t, t+\Delta]$ {\it as soon as}

\begin{eqnarray*}
r^{(i)}(t) > r_0^{(i)}
\end{eqnarray*}
{\it and does not declare otherwise}.

{\it The thresholds are} $r_0^{(i)}=\beta_i/\alpha$, {\it if}

\begin{eqnarray}\label{qq6}
\varphi (n,\tau) = \alpha \Lambda n + <\beta, \tau>.
\label{qq6}
\end{eqnarray}

{\it For the general case of} $\varphi (n,\tau )$,
{\it we consider the level} $c$ {\it such that the surface}
$\varphi (n,\tau )=c$ {\it is tangent to}
$\tilde{\Gamma}$ {\it at a point} $Q$. {\it Then}

\begin{eqnarray}\label{qq7}
r_0^{(i)} = - \Lambda \frac{\partial n}{\partial \tau^i} (Q) = - \Lambda
\frac{\partial \varphi}{\partial \tau^i} \bigg /
\frac{\partial \varphi}{\partial n} (Q),
\label{qq7}
\end{eqnarray}
{\it where} $n=n(\tau^1, \ldots ,\tau^k)$ {\it is the} $\tilde{\Gamma}$
{\it function. Conversely, for any point} $Q=(n,\tau )\in \tilde{\Gamma}$
{\it we can find the loss function} $\varphi (n,\tau )$
{\it for which} $Q$ {\it is optimal, i.e.,}

\begin{eqnarray*}
\varphi (Q) = \inf_{\{ \pi \}} \varphi (n(\pi ),\tau (\pi )),
\end{eqnarray*}
{\it where the strategies} $\pi$ {\it are based on} $I=\{ I(t)\}$.

{\bf Remark.} All components of the optimal strategy are
interconnected due to the data $I(t)$ which are common to subregions
$\{G_i\}$.

{\bf Proof.} Since $(dN(t), I(t), \pi (t))$ is ergodic, the time
average (\ref{qq1}) can be identified with the ensemble average
(over $I(t)$) of the single term in (\ref{qq1}) related to the
interval $(t, t+\Delta)$. The same holds for (\ref{qq2}) because
$N(T)/S \to \Lambda$ as $T\to \infty$. From this it follows that the
globally optimal strategy for (\ref{qq3}) can be derived by
optimizing the decision in every interval $(t, t+\Delta)$. Putting
${\bf 1}=(1,\ldots, 1)$, we have

\begin{eqnarray*}
\nonumber n&=&\lim_{\Delta \to 0}E<{\bf 1}-\pi (t), \Delta
N(t)/\Delta >/\Lambda = \\
\nonumber
&=&\lim_{\Delta \to 0}E\{ E<{\bf 1}-\pi (t), \Delta N(t)/\Delta >
|I(t)\} / \Lambda = \\
\nonumber
&=&E<{\bf 1} - \pi (t), r(t)>/\Lambda = 1 - E<\pi (t), r(t)>/\Lambda, \\
\nonumber
\tau &=&E\pi (t).
\end{eqnarray*}
Here, $\tau, \pi \subset [0,1]^k$. Suppose $\varphi (n,\tau)$ is of the
linear form (\ref{qq6}). Then

\begin{eqnarray}\label{qq8}
\varphi (n,\tau) = \alpha \Lambda + E<\pi (t), \beta - \alpha r(t)>.
\label{qq8}
\end{eqnarray}
The components of $\pi (t)$ take on values
in $[0,1]$. Obviously, (\ref{qq8}) has the least value, when we put

\begin{eqnarray}\label{qq9}
\pi^{(i)}(t)=
\begin{cases}
0,                &\text{$\beta^i - \alpha r^i(t)\ge 0$}\\
1,                &\text{$\beta^i - \alpha r^i(t) < 0$}.
\end{cases}
\label{qq9}
\end{eqnarray}

Suppose now that $\varphi (n,\tau)$ is a nonlinear increasing
function with convex level sets, $\{ \varphi \le a\}$. Then there
exists a level $c$ such that the surface $\varphi (n,\tau)=c$ is
tangent to $\tilde{\Gamma}$ at some point $Q$. By the definition of
$\tilde{\Gamma}$, $c$  is the least value of the goal function given
the predictive information $\{ I(t)\}$. Let us construct a plane
that is tangent to $\tilde{\Gamma}$ at the point $Q$, $an+<b,\tau
>=c$. It separates $\tilde{\Gamma}$ and the surface $\varphi(n,\tau
)=c$, because ${\cal E}$ and $\{ \varphi(n,\tau )\le c\}$ are
convex. Therefore, the minimization of $\varphi$ is equivalent to
the minimization of the linear function $an+(b,\tau )$. The use of
(\ref{qq8}) and (\ref{qq9}) yields (\ref{qq7}). Actually, we have
also proved the final part of the statement, because at any point
$Q\in \tilde{\Gamma}$ there  exists a plane of support to
$\tilde{\Gamma}$. $\diamond$

{\bf Prediction efficiency}. At the research stage of prediction,
the efficiency of a time-space strategy $\pi$ is sometimes
characterized by the quantity $e=1-(n+\tilde{\tau})$, where

\begin{eqnarray}\label{qq10}
\tilde{\tau} = \sum^k_{i=1} \lambda_i \tau^{(i)} /\Lambda
\label{qq10}
\end{eqnarray}
is the rate of space-time alarm measured in terms of the rate of
target events, $\{ \lambda_i\}$. One can suggest some reasons in
favor of this choice of $e$.

{\it First}, $\vert e \vert \le 1$ where $e=0$ for all trivial
strategies, i.e., $(n_{\pi},\tau_{\pi})\in D$, and $e=1$ for the
ideal strategy with zero errors.

{\it Second}, $e=(1-n)-\tilde{\tau}$. In this identity the second
term $\tilde{\tau} =\tilde{\tau}(\pi)$ coincides with the rate of
target events, which can be predicted by chance using the same
space-time alarm characteristics $(\tau^{(1)},\ldots, \tau^{(k)})$
as $\pi$ has. Therefore, $e$ determines the rate of nonrandom
successes of the strategy $\pi$.

{\it Third}, $e=e(\pi)$ is proportional to the Euclidian distance,
$\rho (Q,\tilde{D})$, between $Q=(n,\tau )$ and $\tilde{D}$;
moreover, $e=1$ for the ideal strategy having $(n,\tau )=(0,0)=O$.
Therefore, $e(\pi )=\rho (Q,\tilde{D})/\rho (O,\tilde{D})$, i.e.,
$e$ is the relative distance between $\pi$ and the trivial
strategies set in the coordinates $(n, \tau_1,...,\tau_k)$.

Our interpretation of $e$ does not depend on the space parameter
$k$. This is important for the comparison of predictions, because
the space partition $\{ G_i\}$ is an independent element of a
prediction strategy.

{\it Fourth}, $e$ has  the following additivity property:

\begin{eqnarray*}
e = 1-n-\tilde{\tau} = \sum^k_{i=1} (1-n_i-\tau_i)\lambda_i/\Lambda =
\sum^k_{i=1} e_i\lambda_i/\Lambda,
\end{eqnarray*}
where $(n_i,\tau_i)$ and $e_i=1 -n_i-\tau_i$ are respectively the
errors and the efficiency of $\pi$ in subregion $G_i$. This follows
from (\ref{qq10}) and the relation

\begin{eqnarray*}
n = \sum^k_{i=1} n_i\lambda_i/\Lambda .
\end{eqnarray*}
Thus, $e(\pi)$ is a weighted mean of the efficiencies in subregions
$\{ G_i\}$. The additivity of $e$ holds only for linear functions of
the type $e=an+b\tilde{\tau}+c$ (see Appendix~1 for exact
formulation and proof).

To optimize $e=1-(n+\tilde{\tau})$, we must, in accordance with Statement
1, declare an alarm in $G_i\times \Delta t$, as soon as
the probability gain $(PG)$,
$r^{(i)}_{\xi} /\lambda_i$, exceeds the level~1.
This level is a point of equilibrium of $PG$, therefore,
the alarm which  optimizes $e$ can be unstable in the general
case of $\{ I(t)\}$.

The following example is relevant to the stable situation [Molchan, 2002].

{\it Example~1 (characteristic earthquakes)}. Consider the time
prediction problem in which $I(t)$ is the time $u=t-t_k\ge 0$ that
has elapsed since the last event $t_k$. In that case the optimal
strategy for $e=1-n-\tau$ declares an alarm in the interval $(t_k,
t_{k+1})$ as soon as

\begin{eqnarray*}
mF'(u)/(1-F(u)) > 1, \quad t=t_k+u,
\end{eqnarray*}
where $F$ is the distribution of $\Delta_k =(t_{k+1}-t_k)$ and
$m=E\Delta_k$ [Molchan, 2002]. In many interesting cases $F'/(1-F)$
has at most one extremum in the open interval $(0, \infty)$.
Therefore the optimal alarm in $(t_k,t_{k+1})$ consists at most of
two intervals. It is easy to see that

\begin{eqnarray*}
e = \int_0^\infty [F'(u) - (1-F(u))/m]_+\,du,
\end{eqnarray*}
where $[a]_+=a$, if $a>0$ and $[a]_+=0$, if $a<0$. The following
table presents values of $e$ depending on the coefficient of
variation $V=\sigma /m$ ($\sigma^2$ is the variance of $F$) for
three types of distributions $F$, viz., Weibull $(F(x)=1-\exp
(-\lambda x^\alpha))$, Log-Normal, and Gamma $(F'(x)=cx^{\alpha
-1}\exp (-\lambda x))$:

\begin{center}
\begin{tabular}{c|ccccccc}
\hline
$V$ & \qquad &\,\,\,.25 & \qquad & .50 & \qquad &  .75 & \, \\
\hline
$e$ & \qquad &\,\,\,.52 - .60 & \qquad & .32 - .38 & \qquad & .15 - .22 & \, \\
\hline
\end{tabular}
\end{center}

Here all distributions have the same $m$ and $V$ parameters. Note
that $V\simeq 0.6$ for segments of the San Andreas fault, and that
the model has a direct relation to the prediction of characteristic
earthquakes. Therefore, our example with nontrivial prediction  can
be of interest for comparison with other available prediction
methods.

{\bf Trivial Strategies}. In the time prediction case the trivial
strategies are described by the diagonal $n+\tau =1$ of the square
$[0,1]^2$. The end points $(1, 0)$ and $(0, 1)$ correspond to the
so-called {\it optimistic} and {\it pessimistic} strategies (see
Fig.~1). A pessimist maintains alarm all the time, while an optimist
never uses it. These strategies are remarkable, because in a {\it
regular situation} the points (1,0) and (0,1)  are also the end
points of the curve $\Gamma$, that is, trivial strategies may well
be optimal ones. To understand the regular situation better, we
consider the following counterexample.

{\it Example}~2 ({\it nonregular $\Gamma$}).
Let us consider the following model of target events:

\begin{eqnarray}\label{qq11}
dN(t)/dt = \sum_k \delta(t-t_k) + \sum_k \varepsilon_k \delta(t-t'_k),
\quad t'_k = t_k+1.
\label{qq11}
\end{eqnarray}
Here $t_{k+1}-t_k\ge a>0$ are i.i.d.
random variables with the mean $E(t_{k+1}-t_k)=m$ and
$\{ \varepsilon_k \}$ are independent binary random variables with the
distribution $P(\varepsilon_k =1)=p$, $P(\varepsilon_k =0)=1-p$.
In this model there are two types of target events, viz., main shocks
$\{ t_k\}$ and reshocks $\{ t'_k=t_k+1\}$ that may or may not
occur.

To predict $t_{k+1}$ using $I(t)=\{ t_p: t_p<t\}$ and $t_k<t<t_{k+1}$ {it is
sufficient to declare an
alarm at the moment $t_k+a$ and cancel it after $t=t_{k+1}$.
The reshock $t'_k$ is predicted by short-term alarm at the
moment $t_k+1-0$. Now it is not difficult to see that
the end points $(n,\tau )$ of $\Gamma$ are
$((1+p)^{-1}, 0)$ and $(0, 1-a/m)$. These points
correspond to the regular situation, provided that $p=a=0$. $\diamond$

In the case of space-time prediction, the
trivial strategies are described by the equation $n+\tilde{\tau} =1$,
$0\le n$, $\tau_i\le 1$. All solutions
to that equation are obtained as the convex hull of extreme points
$(n,\tau_i =\varepsilon_i, i=1,\ldots ,k)$, where $\varepsilon_i =0$ or 1,
and $n=1-\tilde{\tau}$.

By definition we are in {\it the regular situation}, if all extreme
points of $\tilde{D}$ belong to $\tilde{\Gamma}$. This is true, if
and only if $I=\{ I(t)\}$ is regular in each subregion $G_i$,
$i=1,\ldots ,k$. In the regular situation, strategies that maintain
a continual alarm in part of the area of interest and no alarm in
its supplement are optimal and trivial at once. This type of
strategies includes Kullback's strategy [Kullback, 1959] ("relative
intensity" in the terminology of Holliday et al. [2005]). The
principle of the strategy is as follows. Suppose we know the
epicenter density of target events, $f(g)$. Find the locations where
$f>c$ and declare a continual alarm there. This strategy is used
during the research stage in order to minimize the alarm space
volume.

In the polemical paper by Marzocchi et al. (2003), the Kullback
strategy is used for comparison with the $M8$ algorithm in the
prediction of $M\simeq 8$ $(M\simeq 7.5)$ earthquakes worldwide.
Note that the Kullback strategy has $n+\tilde{\tau}=1$. Therefore,
the relative predictive potential of the $M8$ algorithm can be
measured by the quantity $e=1-(n+\tilde{\tau})$. To estimate
$\tilde\tau$ in a robust manner, we come to a nontrivial problem: to
what degree can low magnitude seismicity (say, $M=4; 6$) be helpful
in estimating the distribution $f(g)$ (see $\lambda_i/\Lambda$ in
(\ref{qq10}))? The problem is simpler for the case of predicting
$M=7.5; 8$ along the Pacific Belt, because one has to compare
smoothed one-dimensional seismicity distributions along the belt.
This important problem unfortunately remains unexplored.

For the moment one can obtain only a rough estimate for the variability of
$n+\tilde{\tau}$ in the $M8$ case. Denoting by $N(T)$ the number of target
events for the monitoring period $T$, we find that a failure-to-predict
will alter $n$ by the amount $\delta n\approx 1/N$ ($10\%$ in the
prediction of $M=8$). According to [Kossobokov, 2005], $\tilde{\tau}$ in the
prediction of $M=7.5; 8$ varies within $5-10\%$ when $M=4,5,6,7$
is used to estimate the density of target events.
Consequently, the variability of $n+\tilde{\tau}$ for $M=8$ does not exceed
$20\%$.

\section{Prediction versus Forecasting}

According to Statement 1, the prediction problem
considered in its simplest version can be split into two. The one
consists in estimating the conditional rate $r(t,g,M)$ of magnitude
$M$ events in a space-time bin $dg\times dt$, while the other reduces
to choosing a threshold $r_0(g)$ for $r(t,g,M)$. This is an important
conclusion for prediction practice, since the first problem is in
the seismologist's full competence, while the second is at the
option of the customer. At first sight, the seismologist has
merely to focus his efforts on the problem of estimating the risk
function $r(t,g,M)$, i.e., on the {\it forecasting} problem.

In our view forecasting is different from prediction in that it
involves no decisions, and prediction statements are probabilistic
in character, namely, a target event $M$ is expected to occur in the
bin $dg\times dt$ with some probability $P(dg,dt)$. For the
small-bin case, $P(dg,dt)\simeq r(t,g,M)dt\,dg$.

At the present time, forecasting dominates the problem of earthquake
prediction.
Prediction proper came to be viewed as a binary forecasting, where there
is no problem of choosing the thresholds.
This transformation of the original prediction problem
calls for some discussion.

When the information $I(t)$ consists of an earthquake catalog,
the problem of modeling $r(t,g,M)$ is equivalent to constructing a
model of the seismic process in the phase space $(t,g,M)$ in terms
of conditional rate. An example is the self-exciting model
(ETAS as it is called today).

Substitution of forecasting for prediction raises a key question:
what model of $r(t,g,M)$ inspires greater confidence? In prediction,
the information $I(t)$ is chosen and transformed in such a way as to
detect characteristic patterns premonitory to individual target
events. At the research stage, the prediction is thought to be the
better, the smaller the errors $n$, $\tilde{\tau}$, or the
combination $n+\tilde{\tau}$, say.

In forecasting, the goal is hazy; forecasting based on the
conditional rate $r(t,g,M)$ is considered to be the better, the
better is an agreement between the model of $r$ and seismicity
observed during a test period. Target events are rare as a rule,
while premonitory phenomena are weak. For that reason the
contribution of the latter into the fitting of the model of $r$ is
small too. Therefore a "good model" of seismicity will  be
determined mainly by typical seismicity patterns, such as clustering
and aftershocks, regardless of whether they are premonitory or not.
Under these conditions it is difficult to expect that the "good
model" can automatically possess predictive properties in relation
to large earthquakes. Therefore, having formally set up thresholds
for $r(t,g)$, we shall arrive at errors $n,\tilde{\tau}$ that are
close to the diagonal $n+\tilde{\tau}=1$, i.e., will obtain a
misleading "objective proof" that large events are unpredictable.

The ETAS model is often considered to be the
most suitable for description of seismicity
[Ogata, 1999; Kagan and Jackson, 2000].
It is defined in a
form convenient for prediction, in terms of the risk function

\begin{eqnarray}\label{qq12}
r(t,g,M) = \sum_{t-T<t_i<t} U(t,g,M\, \vert \,t_i,g_i,M_i) + U_0(g,M)
\label{qq12}
\end{eqnarray}
Here, $U\ge 0$ is the conditional rate of first-generation aftershocks
for an event $(t_i,g_i,M_i)$, and $U_0\ge 0$ is the rate of main shocks.
The parameterization of $U$ and $U_0$ used in (\ref{qq12}) is too
simplistic for prediction purposes.

The ETAS model satisfactorily incorporates the clustering of events,
hence it is convenient for describing aftershocks. It is known that
some target events were preceded by patterns like seismicity
increase and quiescence. When a threshold $r>r_0$ is defined, the
model (\ref{qq12}) will respond to seismicity increase, but not to
quiescence. The values of $r$ are small in quiescent areas. Ogata
[1988] tried to adapt (\ref{qq12}) to deal with prediction of large
events. In order to be able to respond both to seismicity increases
and to quiescence, alarms were to be declared in two cases, when $r$
was large and when $r$ was small enough. This contradicts
Statement~1. The use of two thresholds instead of a single one means
that  (\ref{qq12}) is not the risk function for large events.

It thus appears that prediction of rare events need not rely on a
detailed seismicity model. This can be seen from Example~1, when it
is compared with results of the $M8$ method, as well as from
Statement~1, which asserts that detailed knowledge of
$r(t,g)/\lambda (g)$ is only needed about a fixed level $c=1$. On
the other hand, overfine detail in $r/\lambda$ close to $c=1$ may
inflate the number of false alarms. Considering forecasting instead
of prediction, we change the original goals and may misrepresent the
predictability of rare events.

\section{Predictability and Scale Invariance}

Scaling laws are well known for seismicity: the distribution of
events over energy (the Gutenberg-Richter law), the decay of
seismicity in time following a large earthquake (the Omori law), the
relation between source dimensions and earthquake energy, and
spatial fractality of seismicity. The above list is being rapidly
supplemented in recent years by laws that use scaling over different
combinations of time, space, and energy. An example is the unified
Bak law for the interevent time in a square of size L [Bak et al.,
2002; Molchan and Kronrod, 2007]. Similarity ideas are actively used
in the passage from the prediction of magnitude $M$ to that of
$M-\Delta$. The first attempt in this direction was for the CN
algorithm (see, e.g., [Keilis-Borok and Rotwain, 1990]).

In the ideal case, if seismicity is strictly similar in the phase
space $(t,g,M)$,  the same predictability should be expected for $M$
and $M-\Delta$. In particular, the events with $M$ and $M-\Delta$
are predictable or unpredictable at the same time based on the
$(t,g,M)$ data. The long-continued monitoring of target events using
the M8 algorithm gives the following results [Kossobokov, 2005]: for
the period 1985-2003 the error statistic $n+\tilde{\tau}$ is equal
to $2/11+0.33\simeq 0.5$ and $22/52+0.34\simeq 0.8$, for $M\simeq 8$
and $M\simeq 7.5$, respectively. The difference in $n+\tilde{\tau}$
is substantial. If the difference is statistically significant, then
it is natural to ascribe it to a violation of the similarity
conditions. Indeed, the similarity condition for earthquakes is
changed, when the source dimension is comparable with the width (W)
of the seismogenic lithosphere [Scholz, 1990; Pacheco et al., 1992;
Okal and Romanowicz, 1994]. The $M=7.5, 8.0$ events fall in this
category. Because $W$ is subject to scatter worldwide, the
finite-depth effect must be more relevant to $M=8$ events. There
exist models for which one can neatly identify the size effect and
its relation to predictability. Shapoval and Shnirman [2006]
considered an avalanche model of the Bak type to show that events
whose size is comparable with the size of the system are predictable
similarly to the $M=8$ events in the $M8$ algorithm, i.e.,
$n+\tilde{\tau}\simeq 0.5$. At the same time, the events that obey
the power law distribution over energy are predicted much worse.

Whether the similarity conditions are violated is frequently
inferred from the presence of a bend in the Gutenberg-Richter
frequency-magnitude relation. It is rather difficult to detect
such a bend, especially in a regional environment. In that
context we give a very simple example in order to
demonstrate that the linearity of the frequency-magnitude
relation does not preclude the predictability of individual
magnitudes.

{\it Example}~3 ({\it predictability vs. GR law}). Consider a region
where events with, say, $M=3, 4, 5$, and 6 occur. The $M=3, 4$, and
6 events are mutually independent in space-time. For the sake of
simplicity we assume the distributions of all events to be uniform.
Select $10\%$ of the area, $G$, and require that each $M=5$ event in
$G$ be necessarily followed by a $M=6$ event during a time $\delta$
(the location is left unspecified). This pattern allows the times of
$M=6$ to be predicted based on the $M=5$ events. The prediction
quality depends on the choice of $\delta$. At the same time, the
frequency-magnitude law will hold in the entire area, if the rates
for $M=3, 4$, and 5 are $\lambda (M)=a\cdot 10^{-M}$. This relation
is also true for $M=6$, because one has

\begin{eqnarray*}
\lambda (M=6) = \lambda (M=5)\cdot 10^{-1} = a\cdot 10^{-6}.
\end{eqnarray*}
by construction. The model has an obvious extension to the
space-time prediction. Now since the $M=3, 4$ and $M=5$ events are
independent, it follows that the $M=5$ events are unpredictable. The
result is that, even though the Gutenberg- Richter law holds, only
the $M=6$ events are predictable. This demonstrates that a violation
of the similarity conditions need not entail changes in the
Gutenberg-Richter law.

\section{Conclusion}

\quad 1. The simplest optimization problem of predicting the time of
large events has been extended to the case of space-time prediction.
We have found an analogue of the error diagram and described the
optimal prediction strategies. The possibility and simplicity of
this extension are due to a special choice of the class of goal
functions (see (\ref{qq3})). In this particular case the globally
optimal strategy can be constructed as a combination of locally
optimal decisions. The situation becomes radically different, when
the goal function is not a function of $(n,\tau)$ alone.

2. The optimal prediction is split into two formally independent
problems: modeling of the risk function $r(t,g,M)$ and choosing its
threshold. However, a separate solution of these problems is a
questionable way to real prediction.

3. In the theory presented here, the
volume of space-time alarm $A$ should be measured by
the expected number of target
events rather than geometrically as the product of area and time.
Due to the simple statistical and geometric interpretation of
$e=1-n-\tilde{\tau}$, this quantity is a natural candidate to
represent the prediction efficiency at the
research stage.

4. We demonstrate on an example  that scaling laws in general
do not exclude predictability of events of different magnitudes.

\bigskip

\noindent {\bf Acknowledgments}

\bigskip

This work was supported by the European Commission's Project
12975 (NEST) "Extreme Events: Causes and Consequences (E2-C2)"
and in part by the Russian Foundation for Basic Research.

\bigskip

\newpage

\noindent {\bf R~E~F~E~R~E~N~C~E~S}

\bigskip

\begin{description}

\item
Bak,~P., Christensen,~K., Danon,~L., Scanlon,~T., 2002.
Unified Scaling Law for Earthquakes, {\it Phys. Rev. E}, {\bf 69,} 066106.

\item
Ellis,~S.P., 1985. On optimal statistical decision rule for calling
earthquake alerts. {\it Earthquake prediction Res.} {\bf 3}, 1-10.

\item
Holliday,~J.R., Nanjo,~K.Z., Tiampo,~K.F., Rundle,~J.B., Turcotte,~D.L.,
2005. Earthquake forecasting and its verification,
{\it Nonlinear Processes in Geophysics, arXiv: cond-mat/} 0508476.

\item
Kagan,~Y.Y. \& Jackson,~D.D., 2000. Probabilistic Forecasting of Earthquakes,
(Leon Knopoff's Festschrift), {\it Geophys. J. Int.} {\bf 143}, 438-453.

\item
Kossobokov,~V.G., 2005. Earthquake Prediction: Principles, Implementation,
Perspectives, in V.I.~Keilis-Borok \& A.A.~Soloviev, Eds.
{\it Computational Seismology}, Iss. {\bf 36-1}, 3-175, GEOS.

\item
Kossobokov,~V. \& Shebalin,~P., 2002. Earthquake Prediction,
in V.I.~Keilis-Borok \& A.A.~Soloviev (eds), {\it Nonlinear Dynamics
of the Lithosphere and Earthquake Prediction}, Springer, p.141-207.

\item
Keilis-Borok,~V.I. \& Rotwain,~I.M., 1990. Diagnosis of times of increased
probability of strong earthquakes in different regions of the world:
algorithm CN, {\it Phys. Earth. Planet. Inter.}, {\bf 61}, 57-72.

\item
Keilis-Borok,~V.I., Shebalin,~P., Gabrielov,~A., Turcotte,~D., 2004.
Reverse Tracing of Short-term Earthquake Precursors,
{\it Phys. Earth. Planet. Inter.}, {\bf 145}, 75-85.

\item
Kullback,~S., 1959. {\it Information Theory and Statistics},
J.Wiley \& Sons.

\item
Lehmann,~E.L., 1959. {\it Testing Statistical Hypotheses},
New York. J.Wiley \& Sons.

\item
Lindgren,~G., 1985. Optimal Prediction of Level Crossings in Gaussian
Processes and Sequences. {\it Annals of Probability}, {\bf 13}:3,
804-824.

\item
Marzocchi,~W., Sandri,~L., Boschi,~E., 2003.
On the Validation of Earthquake-forecasting Models: the Case of Pattern
Recognition Algorithms, {\it Bull. Seism. Soc. Am.}, {\bf 93}, 5, 1994-2004.

\item
Molchan,~G.M., 1991. Structure of Optimal Strategies of
Earthquake Prediction, {\it Tectonophysics}, {\bf 193}, 267-276.

\item
Molchan,~G.M., 1997. Earthquake Prediction as a Decision Making Problem,
{\it Pure Appl. Geophys.}, {\bf 149}, 233-247.

\item
Molchan,~G.M., 2002. Earthquake Prediction Strategies: a Theoretical Analysis.
In V.I.~Keilis-Borok \& A.A.~Soloviev (eds),
{\it Nonlinear dynamics of the Lithosphere and Earthquake Prediction},
Springer, p.209-237.

\item
Molchan,~G.M. \& Kagan,~Y.Y., 1992. Earthquake Prediction and its
Optimization, {\it J. Geophys. Res.}, {\bf 97}, 4823-4838.

\item
Molchan,~G.M. \& Kronrod,~T.L., 2007. Seismic Interevent Time: A Spatial
Scaling and Multifractality, {\it Pure Appl. Geophys.}, {\bf 164}, 75-96.

\item
Ogata,~Y., 1988. Statistical Models for Earthquake Occurrences and Residual
Analysis for Point Processes, {\it J. Amer. Stat. Assosiation},
{\bf 83}, 401, 9-27.

\item
Ogata,~Y., 1999. Seismicity Analysis Through Point-processes Modeling:
A Review, {\it Pure Appl. Geophys.}, {\bf 155}, 471-457.

\item
Okal,~E.A. \& Romanowicz, B., 1994. On the Variation of $b$-values with
Earthquake Size, {\it Phys. Earth. Planet. Inter.}, {\bf 87}, 55-76.

\item
Pacheco,~J.F., Scholz,~C.H., Sykes,~L.R., 1992.
Changes in Frequency-size Relationship from Small to Large Earthquakes,
{\it Nature}, {\bf 355}, 71-73.

\item
Shapoval,~A.B. \& Shnirman,~M.G., 2006.
How Size of Target Avalanches Influences on Prediction Efficiency.
{\it Int. J. of Modern Phys.}, {\bf 17}, 12, 1777-1790.

\item
Scholz,~C.H., 1990. {\it The Mechanics of Earthquake and Faulting},
Cambridge Univ. Press, New York.

\item
Swets,~J.A., 1973. The Relative Operating Characteristic in Psychology,
{\it Science}, {\bf 182}, 4116, 990-1000.
\end{description}

\newpage
{\bf Appendix~1}

\bigskip

The efficiency $e=1-n-\tilde{\tau}$ belongs to the following class
of continuous functions $f(z)$, $z=(x,y)$: for any $m$ and $p=(p_1,
\ldots p_m)$, $\sum p_i=1$, $0\le p_i\le 1$ there exist such $\{
a_i(p), i=1,\ldots ,m\}$ that

\begin{eqnarray*}
\qquad \qquad \qquad f(\sum^m_1p_iz_i) = \sum f(z_i)a_i(p).
\qquad \qquad \qquad \qquad \qquad \qquad (A1)
\end{eqnarray*}
Here  $z_i=(n_i,\tau_i)$ are errors relevant to the subregion $G_i$,
$p_i=\lambda_i/\Lambda$, and
$\sum\limits^m_1p_iz_i=(n,\tilde{\tau})$.

Let us prove  that any continuous function $f$ with the property
(A1) is linear, i.e., $f(x,y)=ax+by+c$.

It is enough to consider the case $m=2$. One has

\begin{eqnarray*}
\qquad \qquad f(pz_1+qz_2) = f(z_1)a(p)+f(z_2)b(p),\quad q=1-p.\qquad
\qquad \quad (A2)
\end{eqnarray*}
If $f(z_0)\ne 0$, then using limit $z_i\to z_0$ one has

\begin{eqnarray*}
\qquad \qquad \qquad \qquad a(p)+b(p)=1. \qquad \qquad \qquad \qquad \qquad
\qquad \qquad \qquad  (A3)
\end{eqnarray*}
Applying (A2) with $p=q=1/2$ to all $z_1:|z_1-z_0|=R$, $z_2=2z_0-z_1$ and
using (A3), we get

\begin{eqnarray*}
f(z_0) = \int\limits_{|z-z_0|=R}f(z)\,ds.
\end{eqnarray*}
Thus, $f$ is a harmonic function; in particular, $f$ is smooth.

Substitute $z_1=z_0-kqz$, $z_2=z_0-kpz$ in (A2) and differentiate
(A2) with respect to $k$ at $k=0$. Then we get

\begin{eqnarray*}
0 = (-qa+pb)\cdot \rho ,
\end{eqnarray*}
where $\rho =f'_n(z_0)x + f'_{\tau}(z_0)y$, $z=(x,y)$. If $\rho
\ne0$, we have $b=q$ and $a=p$. By (A1), $f$ is linear.

\end{document}